\documentclass[12pt]{article}
\usepackage{a4}
\usepackage{amsfonts}
\usepackage{amsmath,amssymb}
\usepackage{graphicx}
\usepackage{slashed}
\usepackage{amsmath}

\usepackage[english]{babel}

\usepackage{multirow}
\usepackage[margin=0.8in]{geometry}

\makeatother

\def\un#1{\relax\ifmmode\@@underline#1\else
        $\@@underline{\hbox{#1}}$\relax\fi}

\let\du=\du                     

\def\a{\alpha}
\def\b{\beta}
\def\c{\chi}
\def\d{\delta}

\def\g{\gamma}
\def\h{\eta}

\def\j{\psi}

\def\l{\lambda}
\def\m{\mu}
\def\n{\nu}

\def\p{\pi}

\def\r{\rho}
\def\s{\sigma}

\def\x{\xi}

\def\D{\Delta}

\def\G{\Gamma}

\def\L{\Lambda}

\def\S{\Sigma}

\def\ve{\varepsilon}

\def\cv{{\cal V}}

\def\bo{{\raise-.3ex\hbox{\large$\Box$}}}               
\def\pa{\partial}                                       
\def\TH{{\raise.2ex\hbox{$\displaystyle \bigodot$}\mskip-4.7mu \llap H \;}}
\def\face{{\raise.2ex\hbox{$\displaystyle \bigodot$}\mskip-2.2mu \llap {$\ddot
        \smile$}}}                                      

\def\Bar#1{\overline{#1}}                       
\def\VEV#1{\left\langle #1\right\rangle}        
\def\leftrightarrowfill{$\mathsurround=0pt \mathord\leftarrow \mkern-6mu
        \cleaders\hbox{$\mkern-2mu \mathord- \mkern-2mu$}\hfill
        \mkern-6mu \mathord\rightarrow$}
\def\dvec#1{\vbox{\ialign{##\crcr
        \leftrightarrowfill\crcr\noalign{\kern-1pt\nointerlineskip}
        $\hfil\displaystyle{#1}\hfil$\crcr}}}           
\def\dt#1{{\buildrel {\hbox{\LARGE .}} \over {#1}}}     

\def\frac#1#2{{\textstyle{#1\over\vphantom2\smash{\raise.20ex
        \hbox{$\scriptstyle{#2}$}}}}}                   
\def\sfrac#1#2{{\vphantom1\smash{\lower.5ex\hbox{\small$#1$}}\over
        \vphantom1\smash{\raise.4ex\hbox{\small$#2$}}}} 
\def\bfrac#1#2{{\vphantom1\smash{\lower.5ex\hbox{$#1$}}\over
        \vphantom1\smash{\raise.3ex\hbox{$#2$}}}}       
\def\afrac#1#2{{\vphantom1\smash{\lower.5ex\hbox{$#1$}}\over#2}}    

\def\[{\lfloor{\hskip 0.35pt}\!\!\!\lceil}
\def\]{\rfloor{\hskip 0.35pt}\!\!\!\rceil}
\def\Lag{{\cal L}}
\def\du#1#2{_{#1}{}^{#2}}

\def\fracm#1#2{\hbox{\large{${\frac{{#1}}{{#2}}}$}}}
\def\ha{{\fracmm12}}

\def\Tr{{\rm Tr}}

\def\un{\underline}
\def\fracmm#1#2{{{#1}\over{#2}}}

\def\low#1{{\raise -3pt\hbox{${\hskip 0.75pt}\!_{#1}$}}}

\def\Dot#1{\buildrel{_{_{\hskip 0.01in}\bullet}}\over{#1}}
\def\dt#1{\Dot{#1}}

\newskip\humongous \humongous=0pt plus 1000pt minus 1000pt
\def\caja{\mathsurround=0pt}
\def\eqalign#1{\,\vcenter{\openup2\jot \caja
        \ialign{\strut \hfil$\displaystyle{##}$&$
        \displaystyle{{}##}$\hfil\crcr#1\crcr}}\,}
\newif\ifdtup

\def\({\left(}
\def\){\right)}

\def\beq{\begin{equation}}
\def\eeq{\end{equation}}
\def\bea{\begin{eqnarray}}
\def\eea{\end{eqnarray}}

\newcommand{\be}{\begin{equation}}
\newcommand{\ee}{\end{equation}}
\newcommand{\nbe}{\begin{equation*}}
\newcommand{\nee}{\end{equation*}}

\newcommand{\lb}{\label}

\begin{document}

\thispagestyle{empty}

{\hbox to\hsize{
\vbox{\noindent November 2019 \hfill IPMU19-0056}}}
{\hbox to\hsize{
\vbox{\noindent  revised version }}}
\noindent
\vskip2.0cm
\begin{center}

{\Large\bf Gravitino condensate in $N=1$ supergravity coupled 
\vglue.1in to the $N=1$ supersymmetric  Born-Infeld theory }

\vglue.3in

Ryotaro Ishikawa~${}^{a}$ and Sergei V. Ketov~${}^{a,b,c,d}$ 
\vglue.1in

${}^a$~Department of Physics, Tokyo Metropolitan University, \\
1-1 Minami-ohsawa, Hachioji-shi, Tokyo 192-0397, Japan\\
${}^b$~Max Planck Institute for Gravitational Physics (Albert Einstein Institute),\\
 Science Park Potsdam-Golm, Am M\"uhlenberg 1, D-14476, Potsdam-Golm, Germany\\
${}^c$~Research School of High-Energy Physics, Tomsk Polytechnic University,\\
2a Lenin Avenue, Tomsk 634050, Russian Federation \\
${}^d$~Kavli Institute for the Physics and Mathematics of the Universe (WPI),
\\The University of Tokyo Institutes for Advanced Study, Kashiwa, Chiba 277-8583, Japan

\vglue.1in
ishikawa-ryotaro@ed.tmu.ac.jp, ketov@tmu.ac.jp
\end{center}

\vglue.3in

\begin{center}
{\Large\bf Abstract}
\end{center}
\vglue.2in

\noindent The $N=1$ supersymmetric Born-Infeld theory coupled to $N=1$ supergravity in four
spacetime dimensions is studied in the presence of a cosmological term with spontaneous supersymmetry breaking. The consistency is achieved by compensating a negative contribution
to the cosmological term from the Born-Infeld theory by a positive contribution originating from the gravitino condensate. This leads to an identification of the Born-Infeld scale with the supersymmetry breaking scale. The dynamical formation of the gravitino condensate in supergravity is reconsidered and the induced  one-loop effective potential is derived. Slow roll cosmological inflation with the gravitino condensate  as the inflaton (near the maximum of the effective potential) is viable against the Planck 2018 data and can lead to the inflationary (Hubble) scale as high as $10^{12}$ GeV.  Uplifting the Minkowski vacuum (after inflation) to a de Sitter vacuum (dark energy) is possible by the use of the alternative Fayet-Iliopoulos  term. Some major physical consequences of our scenario to reheating are briefly discussed also. 


\newpage

\section{Introduction}

The gravitino condensate and the gravitino mass gap in $N=1$ supergravity \cite{Wess:1992cp} coupled to the Volkov-Akulov field \cite{Volkov:1973ix} in four spacetime dimensions arise as the one-loop effect due to the quartic gravitino interaction coming from  the gravitino contribution to the spacetime (con)torsion \cite{Jasinschi:1983wr,Jasinschi:1984cx}. This is similar to the Nambu-Jona-Lasinio model \cite{Nambu:1961tp} of the dynamical generation of electron mass and the formation of Cooper pairs near the Fermi surface in superconductivity. The dynamical gravitino mass also leads to a positive contribution to the vacuum energy and, hence,  the dynamical supersymmetry breaking too \cite{Deser:1977uq}. Given the standard (reduced) Planck mass as the only (dimensional) coupling constant, the gravitino mass gap should be of the order of Planck scale also, which prevents phenomenological applications of the gravitino condensate to physics under the Planck scale.

However, the effective scale of quantum gravity may be considerably lower than its standard value associated with the (reduced) Planck mass $M_{\rm Pl}=1/\sqrt{8\p G_{\rm N}}\approx 2.4\times 10^{18}$ GeV. It may happen because the effective strength of gravity can depend upon either large or warped extra dimensions in braneworld, or the dilaton expectation value in string theory, or both these factors together \cite{ArkaniHamed:1998rs,Antoniadis:1998ig,Randall:1999ee}.~\footnote{The effective Planck scale may also be dynamically generated \cite{Kubo:2018kho}.}
 The negative results of the Large Hadron Collider (LHC) searches for copious production of black holes  imply that the low-scale gravity models may have to be replaced by the high-scale gravity (or supergravity) models, whose effective Planck scale $\tilde{M}_{\rm Pl}$ is much higher the TeV scale but is still under the standard scale $M_{\rm Pl}$, i.e. 
\be \lb{epl}
1~{\rm TeV} \ll \tilde{M}_{\rm Pl}\ll M_{\rm Pl}~.
\ee
This can be of particular importance to the early Universe cosmology, where the Newtonian limit
does not apply, as well as for high-energy particle physics well above the electro-weak scale.

In this scenario, supergravity may play the crucial role in the description of cosmological inflation, reheating, dark energy and dark matter, see e.g., Ref.~\cite{Ketov:2019mfc} and the references therein. For instance, it is unknown which physical degrees of freedom were present during inflation, while supergravity may be the answer. Describing inflation and a positive cosmological constant (dark energy) in supergravity is non-trivial, especially when one insists on the minimalistic hidden sector. Inflation is driven by positive energy so that it breaks supersymmetry (SUSY) spontaneously. As a (model-independent) consequence, goldstino should be present during inflation in supergravity cosmology. The goldstino effective action is {\it universal} and is given by the Akulov-Volkov (AV) action up to field redefinition \cite{Hatanaka:2003cr,Kuzenko:2009ym}.  As was demonstrated in Refs.~\cite{Abe:2018plc,Abe:2018rnu}, the viable description of inflation and dark energy in supergravity can be achieved by employing an $N=1$ {\it vector} multiplet with its
N=1 supersymmetric Born-Infeld (BI) action \cite{Cecotti:1986gb} in the presence of the alternative Fayet-Iliopoulos (FI) term \cite{Fayet:1974jb,Cribiori:2017laj,Kuzenko:2018jlz,Aldabergenov:2018nzd,Cribiori:2018dlc}  without gauging the R-symmetry.~\footnote{In Refs.~\cite{Aldabergenov:2016dcu,Aldabergenov:2017bjt,Addazi:2017ulg}, the $N=1$ massive vector multiplet, unifying Starobinsky inflaton (scalaron) \cite{Starobinsky:1980te} and goldstino (photino)  was used together with the BI action, the FI term, the chiral (Polonyi) multiplet representing the hidden SUSY breaking sector, and the massive gravitino as the Lightest SUSY Particle (LSP) for dark matter.}

In this paper we also employ an $N=1$ vector multiplet with its $N=1$ supersymmetric BI action that automatically contains the goldstino  (AV) action, but we choose the gravitino condensate as the inflaton. A dynamical  SUSY breaking is achieved at the very high scale with the vanishing cosmological constant. The extra (FI) mechanism of spontaneous SUSY breaking is then used to uplift a Minkowski vacuum to a de Sitter (dS) vacuum.  

The BI theory has solid motivation. It is expected that Maxwell electrodynamics does not remain unchanged  up to the Planck scale, because of its internal problems related to the Coulomb singularity and the unlimited values of electro-magnetic field. This motivated Born and Infeld \cite{Born:1934gh} to propose the non-linear vacuum electrodynamics with the Lagrangian (in flat spacetime)
\be \lb{binf}
\Lag_{\rm BI} = -M_{\rm BI}^4\sqrt{-\det\left(\h_{\m\n}+M_{\rm BI}^{-2}F_{\m\n}\right)}=-M_{\rm BI}^4
-\frac{1}{4}F^2 +{\cal O}(F^4)~,
\ee
where $\h_{\m\n}$ is Minkowski metric, $F_{\m\n}=\pa_{\m}A_{\n} -\pa_{\n}A_{\m}$, and
$F^2=F^{\m\n}F_{\m\n}$.  The constant term on the right-hand-side of Eq.~(\ref{binf}) can be ignored in flat spacetime. 
 The BI theory has the new scale $M_{\rm BI}$ whose value cannot exceed the GUT scale where electro-magnetic interactions merge with strong and weak interactions. On the other hand, we need $M_{\rm BI}<\tilde{M}_{\rm Pl}$ in order to ignore quantum gravity corrections. The BI theory naturally emerges (i) in the bosonic part of the open superstring effective action \cite{Fradkin:1985qd}, (ii) as part of Dirac-Born-Infeld (DBI) effective action of a D3-brane \cite{Leigh:1989jq}, and (iii) as part of Maxwell-Goldstone action describing partial supersymmetry breaking of $N=2$ supersymmetry to $N=1$ supersymmetry \cite{Bagger:1996wp,Rocek:1997hi}.~\footnote{See Refs.~\cite{Ketov:1998ku,Ketov:1998sx,Bellucci:2000ft,Bellucci:2012nz} for the extensions of BI theory to extended supersymmetry and higher dimensions.}  The peculiar non-linear structure of the BI theory is responsible for its electric-magnetic (Dirac) self-duality, taming the Coulomb self-energy of a point-like electric charge, and causal wave propagation (no shock waves and no superluminal propagation) --- see e.g., Refs.~\cite{Ketov:2001dq,Ketov:1996bm} and the references therein for a review and non-abelian extensions of BI theory. All that adds more reasons for the use of the BI structure.

In a curved spacetime with metric $g_{\m\n}$ the BI action is usually defined as the difference between two spacetime densities,
\be \lb{bia} 
S_{\rm BI,standard} = M_{\rm BI}^4\int d^4x\,\left[ \sqrt{-\det(g_{\m\n})} - \sqrt{-\det\left(g_{\m\n}+
M_{\rm BI}^{-2}F_{\m\n}\right)} \,\right]~,
\ee
where the first term has been added "by hand" in order to eliminate the cosmological constant arising from the second term and in Eq.~(\ref{binf}).  In this paper we propose the gravitino condensation as the origin and the mechanism of such cancellation in the supergravity extension of the BI theory with spontaneously broken SUSY. 

The $N=1$  (rigid) supersymmetric extension of BI theory is also self-dual \cite{Kuzenko:2000tg}. The supersymmetric BI theory coupled to $N=1$ supergravity (i.e., the locally supersymmetric extension of Eq.~(\ref{bia}))  was constructed in Ref.~\cite{Kuzenko:2005wh}. 

Our paper is organized as follows. In Sec.~2 we provide more details how to deal with a cosmological constant and spontaneous supersymmetry breaking in the context of a supersymmetric BI theory coupled to supergravity, and relate the BI scale to the spontaneous SUSY breaking scale. Most comments in Sec.~2 are known in the literature and are recalled to justify consistency of our approach. In Sec.~3 we study the dynamical gravitino condensate arising from the one-loop effective action of pure supergravity, and investigate the induced scalar potential. Slow roll inflation with the gravitino condensate playing the role of inflaton is numerically studied in Sec.~4.  Uplifting the Minkowski vacuum to a de Sitter vacuum by the use of the alternative  FI term is proposed in Sec.~5. Our conclusion is Sec.~6.  We use the supergravity notation of  Ref.~\cite{Wess:1992cp}.
\vglue.2in

\section{Spontaneous SUSY breaking, AV and BI actions, and their coupling to supergravity}

We recall that the AV Lagrangian in flat spacetime is given by \cite{Volkov:1973ix}
\be \lb{avl}
\mathcal{L}_{\rm AV} = -M^4_{\rm susy}\det\left(\d^a_b+\fracmm{i}{2M^4_{\rm susy}}\bar{\l}\g^a\pa_b\l\right) =  -M^4_{\rm susy} -\frac{i}{2}\bar{\l}\g\cdot\pa\l +{\cal O}(\l^4)~,
\ee
where $\l(x)$ is a Majorana fermion field of spin 1/2. This fermionic field is called the {\it goldstino} because the AV action has the spontaneously broken non-linearly realized rigid SUSY under the transformations
\be \lb{avs} 
\d\l= M^2_{\rm susy}\ve +\fracmm{i}{M^2_{\rm susy}}(\bar{\ve}\g^a\l)\pa_a\l
\ee
with the infinitesimal Majorana spinor parameter $\ve$, so that the goldstino is a Nambu-Goldstone fermion indeed. The AV theory (\ref{avl}) has the spontaneous SUSY breaking scale  $M_{\rm susy}$. 

A coupling of the AV action to supergravity is supposed to generate a gravitino mass via the so-called super-Higgs effect \cite{Wess:1992cp} when the gravitino "eats up" the goldstino and thus gets the right number of the physical degrees of freedom. However, it is impossible to couple the AV action to supergravity in the manifestly supersymmetric way (i.e with the linearly realized SUSY) when using standard supermultiplets or unconstrained superfields because of the mismatch in the numbers of the bosonic and fermionic physical degrees of freedom.~\footnote{The manifestly supersymmetric description is, nevertheless, possible at low energies when embedding the goldstino into the constrained chiral superfield $\tilde{X}$ obeying the nilpotency condition $\tilde{X}^2=0$ \cite{Rocek:1978nb,Lindstrom:1979kq,Komargodski:2009rz,Antoniadis:2014oya}. We avoid that goldstino superfield because it is problematic at higher energies and in quantum theory.} We embed goldstino into a standard {\it vector} supermultiplet, i.e. identify goldstino with photino, and use an $N=1$ supersymmetric BI action for the vector multiplet, because it is well motivated at very high energies and includes the goldstino AV action up to a field redefinition \cite{Hatanaka:2003cr,Kuzenko:2009ym}.

The supersymmetric extension of the BI action (\ref{bia}) minimally coupled to supergravity in curved superspace of the (old-minimal) supergravity (in a superconformal gauge) with the vanishing cosmological constant and the vanishing  gravitino mass is given by 
\be \lb{ssbi}  \eqalign{
S_{\rm SBI}[V]= & \frac{1}{4}\left( \int d^4xd^2\theta {\cal E} W^2 + {\rm h.c.}\right) +
\frac{1}{4} M_{\rm BI}^{-4} \int d^4xd^2\theta d^2\bar{\theta} E\fracmm{ W^2{\bar W}^2}{1+\frac{1}{2}A+\sqrt{1+A+\frac{1}{4}B^2}}~~,\cr
& A=\frac{1}{8}M_{\rm BI}^{-4}\left( {\cal D}^2W^2+{\rm h.c.}\right)~,   \quad
B=\frac{1}{8}M_{\rm BI}^{-4}\left( {\cal D}^2W^2-{\rm h.c.}\right)~,   \cr}
\ee
where ${\cal E}$ is the chiral (curved) superspace density, $E$ is the full (curved) superspace density, ${\cal D}^{\a}$ are the covariant spinor derivatives in superspace, $W^{\a}$ is the chiral 
gauge-invariant field strength,
\be \lb{sfs}
W_{\a}=-\fracm{1}{4}\left( \bar{\cal D}^2-4{\cal R}\right){\cal D}_{\a}V~~,
\ee
of the gauge real scalar superfield pre-potential $V$ describing an $N=1$ vector multiplet, 
${\cal R}$ is the chiral (scalar curvature) supergravity superfield, $W^2=W^{\a}W_{\a}$ and  ${\cal D}^2= {\cal D}^{\a} {\cal D}_{\a}$ \cite{Wess:1992cp}. 

The action (\ref{ssbi}) is obtained from the standard (Bagger-Galperin) action 
 \cite{Bagger:1996wp} 
 \be \lb{bga}
 S_{\rm BG}[W,\Bar{W}]=\fracmm{1}{4}\int d^4x d^2\theta\, X +{\rm h.c.}~,\quad
 X + \fracmm{1}{4M^4_{\rm BI}}X\Bar{D}^2\Bar{X}=W^2~~,
 \ee
 in terms of the constrained chiral superfield $X$ after solving the constraint in Eq.~(\ref{bga}) and then
 minimally coupling the resulting action with the supergravity in curved superspace \cite{Ketov:2001dq,Kuzenko:2018jlz}, where spacetime metric $g_{\m\n}$ is replaced by vierbein $e^a_{\mu}$ and is extended to an off-shell  supermultiplet  $(e^a_{\mu}, \j_{\m},M,b_{\mu})$, with $\j_{\mu}$ as the  Majorana gravitino field, whereas the complex scalar $M$ and the real vector field $b_{\mu}$ are the auxiliary fields.~\footnote{The auxiliary fields of the supergravity multiplet do not play the significant role in our investigation and are ignored below.} 
 
 The gauge vector (photon) field $A_{\mu}$  is extended in SUSY to  an off-shell (real) gauge vector multiplet (or a general real superfield) $V$ with the field components
\be  \lb{vm}
V=(C,\c,H,A_{\mu}, \l,D)~,
\ee
where $\l$ is Majorana fermion called the photino, $D$ is the auxiliary field, while the rest of the fields $(C,\c,H)$ are the super-gauge degrees of freedom that are ignored in what follows. 

Disturbing the action (\ref{ssbi}) via adding a negative cosmological constant $-M^4_{\rm BI}$ to restore the original BI action (\ref{binf}) explicitly breaks SUSY that, however, can be restored by modifying the action and the SUSY transformation laws \cite{Deser:1977uq,Cribiori:2018dlc}. As a result, it was found that the deformed (new) BI action cannot have a non-vanishing cosmological constant but can have a spontaneously broken local SUSY with a non-vanishing gravitino mass related to the SUSY breaking scale $M_{\rm BI}$.  This does not explain, however, the physical origin of the compensating positive term $+M^4_{\rm BI}$ needed in particular. We explain its origin by the gravitino condensation (Sec.~3).  To illustrate those features, we add a few simple arguments below.  

In order to cancel the SUSY variation of the cosmological constant multiplied by 
$\sqrt{-\det(g_{\m\n})}=e$  due to $\d_{\rm susy} e^a_{\mu}= -i\tilde{M}^{-1}_{\rm Pl}(\bar{\ve}\g^a\j_{\mu})$ with the infinitesimal SUSY parameter $\ve(x)$, we have to add the photino-gravitino mixing term 
\be \lb{phgr} 
              -ie\fracmm{M^2_{\rm BI}}{\tilde{M}_{\rm Pl}}(\bar{\l}\gamma^{\mu}\j_{\mu})
\ee
to the Lagrangian, and simultaneously demand the supersymmetric variation of the photino $\l$ as
\be \lb{golg}
\d_{\rm susy}  \l = M^{2}_{\rm BI}\ve+\ldots~,
\ee
where the dots stand for the other field-dependent terms. The identification of the photino $\l$ with the
goldstino of the spontaneously broken local SUSY requires  
\be \lb{indet}
M_{\rm BI} = M_{\rm susy}
\ee  
already by comparison of Eqs.~(\ref{avs}) and (\ref{golg}). This may be not surprising after taking into account that the initial (rigid) Bagger-Galperin action (\ref{bga}) has the second (spontaneously broken and non-linearly realized) SUSY whose transformation law is similar to that of Eq.~(\ref{avs}). However, our deformed super-BI action in supergravity does not respect another SUSY by construction.

The SUSY-restoring deformation comes together with the gravitino mass term having the mass parameter $m^2=\frac{1}{3}M^4_{\rm BI}/\tilde{M}_{\rm Pl}^{2}$, and the modification of the gravitino SUSY transformation law as 
$\d_{\rm susy}\j_{\mu}=-2\tilde{M}_{\rm Pl}(D_{\mu}\ve +\frac{1}{2}m\g_{\m})+\ldots$. This also implies (by local SUSY) the presence of the goldstino mass term in the Lagrangian with the same mass parameter $m$  \cite{Deser:1977uq}. Hence, the super-Higgs effect is in place.

The recovery of the AV action from the super-BI action is possible by identifying
the goldstino $\l_{\a}$ with the leading field component of the superfield $W_{\a}$ and projecting 
the other fields out, $F_{\m\n}(A)=D=\j_{\m}=0$ in the absence of gravity, $e^a_{\m}=\d^a_{\m}$. Then
the action (\ref{bga}) reduces to the AV action in Eq.~(\ref{avl}) up to a field redefinition in the higher-order terms --- see Ref.~\cite{Kuzenko:2011tj} for details.  The same conclusions are supported by the superconformal tensor calculus in supergravity \cite{Freedman:2012zz}. In our approach, the AV action is thus the fermionic {\it fragment} of the supersymmetric BI theory coupled to supergravity with the spontaneously broken SUSY at the scale $M_{BI}$. In the next Sec.~3 we concentrate on the pure supergravity sector of our theory, ignoring the gravitino-photino mixing (i.e., taking into consideration only spin-3/2 gravitino components), just for simplicity. Accounting of a spin-1/2 photino contribution is beyond the scope of our investigation in this paper.

\section{One-loop effective action and gravitino condensate}

The classical supergravity Lagrangian $\mathcal{L}_{\rm SUGRA}$ besides the Einstein-Hilbert and Rarita-Schwinger terms,
\be  \lb{eh}
\mathcal{L}_{\rm EH} = 
-\fracmm{\tilde{M}^2_{\rm Pl}}{2}eR
\ee
and
\be \lb{rs}
\mathcal{L}_{\rm RS} =  -
\fracm{1}{2}\ve^{\m\n\l\r}\bar{\j}_{\m}\g_5\g_{\n}D_{\l}\j_{\r}~~,
\ee
respectively, also has the quartic gravitino coupling, 
\be \lb{sugra}
\mathcal{L}_{\rm quartic} = 
\fracm{11}{16}\tilde{M}_{\rm Pl}^{-2}\left[ (\bar{\j}_{\m}\j^{\m})^2 - (\bar{\j}_{\m}\g_5\j^{\m})^2\right] -
\fracm{33}{64} \tilde{M}_{\rm Pl}^{-2}(\bar{\j}^{\m}\g_5\g_{\n}\j^{\m})^2~,
\ee
originating from the spacetime (con)torsion in the covariant derivative of the gravitino field in its kinetic term, in the second-order formalism for supergravity \cite{Wess:1992cp}.~\footnote{We separate the
quartic terms \lb{sugra} from the minimal term in Eq.~(\ref{rs}).}

Since the supergravity action is invariant under the local SUSY, whose gauge field is $\j_{\m}$, one can choose the (physical) gauge condition $\g^{\m}\j_{\m}=0$, which implies $(\bar{\j}_{\m}\S^{\m\n}\j_{\n})= -\ha \bar{\j}_{\m}\j^{\m}$, in the notation $\S_{\m\n}=\frac{1}{4}[\g^{\m},\g^{\n}]_-$, and rewrite the (non-chiral) quartic gravitino term in Eq.~(\ref{sugra}) as
\be \lb{ro}
\mathcal{L}_{\rm quartic} = \sqrt{11} \tilde{M}_{\rm Pl}^{-1}\r (\bar{\j}_{\m}\S^{\m\n}\j_{\n}) - \r^2~,
\ee
where the real scalar field $\r$ has been introduced. As is clear from Eq.~(\ref{ro}), a gravitino condensate leads to the non-vanishing Vacuum Expectation Value (VEV), $\VEV{\r}\equiv \r_0\neq 0$, whereas $\r_0$ contributes to the gravitino mass.

The one-loop contribution to the effective potential $V_{\rm 1-loop}(\r)$ of the scalar field $\r$ together
with its kinetic term arise after quantizing the gravitino sector and taking the Gaussian integral over 
$\j_{\m}$ in the gauge $\g^{\m}\j_{\m}=0$. This yields the one-loop contribution to the quantum effective action in the standard form
\be \lb{loop}
\G_{\rm 1-loop}=-\fracm{i}{2}\Tr \ln \D (\r)~,
\ee
where $\D (\r)$ stands for the kinetic operator in the gravitino action and the interaction with gravity
is ignored $(e^a_{\m}=\d^a_{\m})$. The one-loop contribution to the $\r$-scalar potential (i.e. the terms without the spacetime derivatives in Eq.~(\ref{loop})) was first computed in Refs.~\cite{Jasinschi:1983wr,Jasinschi:1984cx} with the result
\be \lb{lpot}
V_{\rm 1-loop} =\lim_{\cv\to\infty} \left[ \fracmm{-1}{2\cv}\sum_{n=1}^{\infty}
\fracmm{(\sqrt{11}\tilde{M}_{\rm Pl})^{2n}}{2n} \Tr (P_{ab}\r)^{2n}\right]=
-\fracmm{4}{(2\p)^4}\int^{\L} d^4p\, \ln \left( 1 + 11\tilde{M}_{\rm Pl}^{-2}\fracmm{\r^2}{p^2}\right)~,
\ee 
in terms of the standard massless gravitino propagator (in momentum space)
\be \lb{prop}
P_{ab} = - \fracmm{i}{2}\fracmm{\g_b\g^{\m}p_{\m}\g_a}{p^2}~~,
\ee
the spacetime 4-volume regulator $\cv$  and the Ultra-Violet (UV) cutoff $\L$, with the trace $\Tr$ acting on all  variables. 

The one-loop contribution (\ref{loop}) expanded up to the 2nd order in the spacetime
derivatives also yields the $\r$-kinetic term subject to the wave function renormalization 
(i.e. with the $Z$ factor), so that the initially auxiliary scalar field $\r$ becomes dynamical with a
mass $M_c$. The specific calculations can be found in the literature \cite{Jasinschi:1983wr,Jasinschi:1984cx,Buchbinder:1989gi,Ellis:2013zsa,Alexandre:2013iva}, and the effective potential reads~\footnote{The quantum effective action may have the {\it imaginary} part (sometimes lost in perturbation theory) that contributes to decay of the gravitino condensate after inflation. Our considerations are limited to the inflationary era by assuming the scale of the imaginary part to be much less the scale of inflation.} 
\be \lb{gcon1} 
V(\r)\equiv V_{\rm classical}(\r) + V_{\rm 1-loop}(\r)  =   \r^2 -\fracmm{4}{(2\p)^4}\int^{\L} d^4p\, \ln \left( 1 + 11\tilde{M}_{\rm Pl}^{-2}\fracmm{\r^2}{p^2}\right)~~.
\ee 
Our result of taking the four-dimensional integral in Eq.~(\ref{gcon1}) is given by ({\it cf.} Refs.~\cite{Jasinschi:1983wr,Jasinschi:1984cx})
\be \lb{gcon}
V(\r)  =  \r^2 +\fracmm{1}{8\p^2}\left\{  \fracmm{121\r^4}{\tilde{M}_{\rm Pl}^{4}}\ln 
\left(1+\fracmm{\tilde{M}_{\rm Pl}^{2}\Lambda^2}{11\r^2}\right) -
\fracmm{11\r^2\Lambda^2}{\tilde{M}_{\rm Pl}^{2}} -
\L^4\ln\left(1+\fracmm{11\r^2}{\tilde{M}^2_{\rm Pl}\L^2}\right) \right\}~.
\ee

The logarithmic  scaling of the wave function renormalization of $\r$ in the one-loop approximation yields the factor  proportional to   $\ln\left(\fracmm{\L^2}{\m^2}\right)$, where $\mu$ is the renormalization scale. Hence, the canonical (physical) scalar $\phi$ is given by \cite{Ellis:2013zsa} 
\be \lb{logs}
\phi =  {\rm const.} \sqrt{ \ln\left(\fracmm{\L^2}{\m^2}\right) } \tilde{M}_{\rm Pl}^{-1}\r \equiv \tilde{w} M_{\rm Pl}\s~,
\ee
where we have introduced the dimensionless (renormalization) constant $\tilde{w}$ as the parameter. We also use the other dimensionless quantities 
\be \lb{dless}
\s = \tilde{M}_{\rm Pl}^{-2} \r~, \quad \tilde{M}_{\rm Pl}^{-1} \Lambda =\tilde{\L} \quad {\rm and} \quad \tilde{M}_{\rm Pl}^{-1}  M_{\rm BI}=\a~,
\ee
which allow us to rewrite the {\it full\/} scalar potential  as
\be \lb{pot2}
V(\s)\tilde{M}_{\rm Pl}^{-4} =\s^2 - \fracmm{1}{8\p^2} \left\{  \tilde{\L}^4 \ln \left(1+
\fracmm{11\s^2}{\tilde{\L}^2}\right) -121\s^4\ln \left(1+ \fracmm{\tilde{\L}^2}{11\s^2}\right) + 11\s^2\tilde{\L}^2 \right\}+\a^4~,
\ee
where we have added the contribution of the first term on the right-hand-side of Eq.~(\ref{binf}).

The scalar potential (\ref{pot2})  has the double-well shape and is bounded from below, see Fig.~1, provided that
\be \lb{bound}
 \tilde{\L}^2 > \fracmm{4\p^2}{11}\approx 3.59~, \quad {\rm or} \quad \tilde{\L} >  \fracmm{2\p}{\sqrt{11}}\approx 1.89~~.
\ee 
There is a local maximum  at $\r=\s=0$ with the positive height $M^4_{\rm BI}$. A similar potential near its maximum was used for describing slow-roll inflation with the inflaton field $\phi$ \cite{Ellis:2013zsa}, see the next Sec.~4 for more. There are also two stable Minkowski vacua at $\r_c\neq 0$.

\begin{figure}[t]
\begin{center}
\vspace{1cm}
\includegraphics[width=15cm,height=8cm]{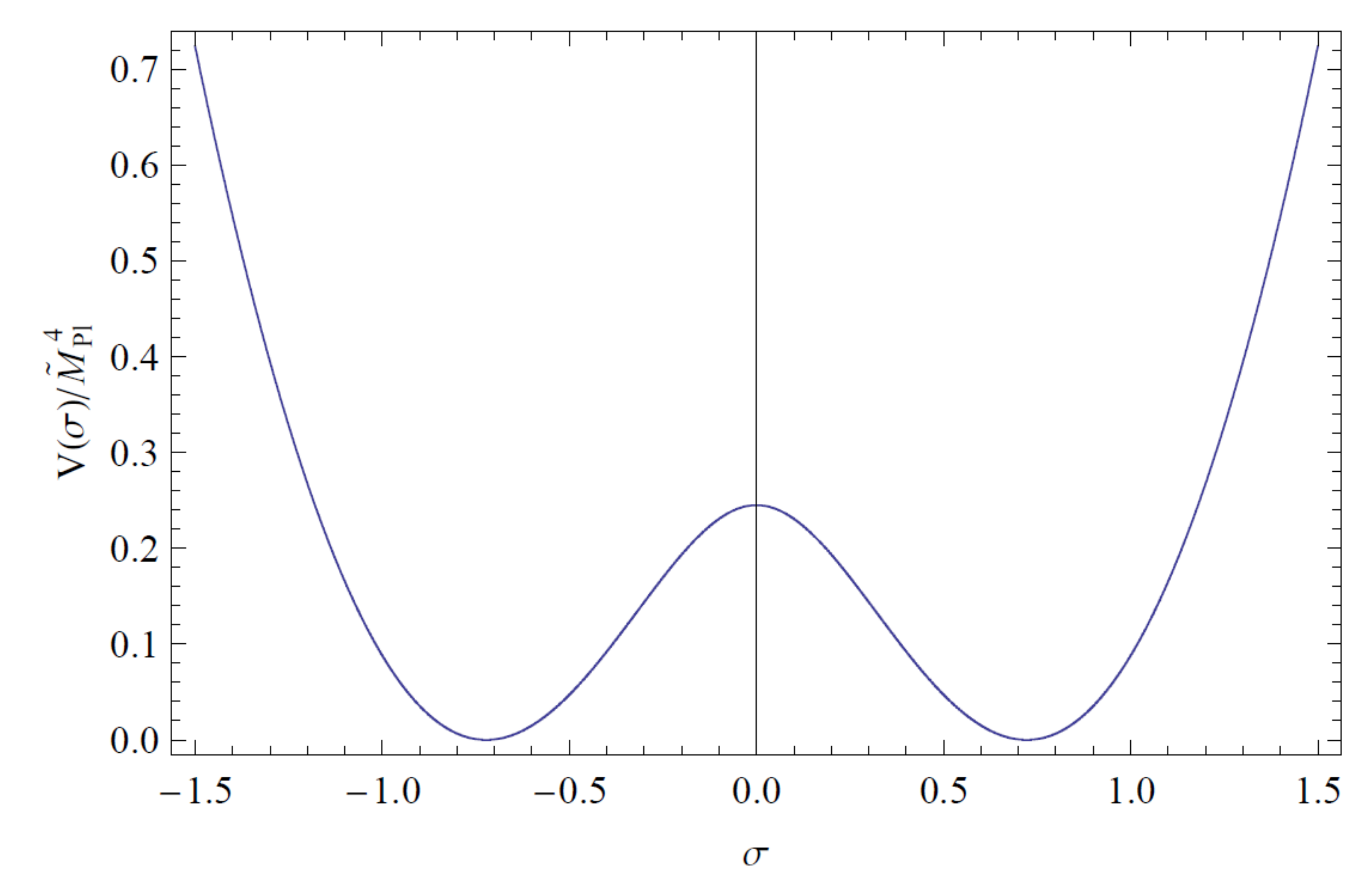}
\caption{The profile of the $V(\sigma)$ function in Eq.~(\ref{pot2}). } \label{fig:1}
\end{center}
\end{figure}

According to the previous Section, supersymmetry requires the scalar potential (\ref{pot2}) to vanish at the minimum, i.e. $V(\sigma_c)=0$. In addition, according to Eq.~(\ref{ro}), the $\r_c\neq 0$ determines the gravitino condensate mass 
\be \lb{gcm} 
m_{\phi}=\sqrt{11}\r_c/\tilde{M}_{\rm Pl}=\sqrt{11}\tilde{M}_{\rm Pl}\s_c. 
\ee

The non-vanishing values of $\r_c$ and $\s_c$ are determined by the condition $dV/d (\s^2)=0$ that yields a transcendental equation,
\be \lb{crit}
121\s^2_c \ln \left( 1+\fracmm{\tilde{\L}^2}{11\s_c^2}\right) = 11 \tilde{\L}^2 -4\p^2 >0~.
\ee

The hierarchy between the inflationary scale $H_{\rm inf.}$, the BI scale $M_{\rm BI}$, the SUSY breaking scale 
$M_{\rm susy}$, the (super)GUT scale $M_{\rm GUT}$, the effective gravitational scale $\tilde{M}_{\rm Pl}$ and the Planck scale $M_{\rm Pl}$ in our approach reads
\be \lb{hier1}
H_{\rm inf.} \ll M_{\rm BI}= M_{\rm susy} \approx M_{\rm GUT} \approx \tilde{M}_{\rm Pl} \ll 
M_{\rm Pl}~~,
\ee
where "much less" means the 2-3 orders of magnitude "less" (in GeV), and "approximately" means the same order or magnitude, see the next Section for our numerical estimates. As regards the GUT scale, we take $M_{\rm GUT}\approx {\cal O}(10^{15})$ GeV.

\section{Gravitino condensate as inflaton}

A slow-roll inflation induced by gravitino condensation in supergravity was proposed and studied by Ellis and Mavromatos in Ref.~\cite{Ellis:2013zsa}. Since our induced scalar potential differs from that of Ref.~\cite{Ellis:2013zsa}, we reconsider this inflation in this Section, by using $\tilde{\L}$ 
and $\tilde{w}$ as the phenomenologically adjustable parameters.

A slow roll is possible near the maximum of the scalar potential (\ref{pot2}). Since the height of the
potential at the maximum is related to the inflationary Hubble scale $H_{\rm inf.}$ by Friedmann equation,
\be \lb{fried}
V_{\rm max.} = 3M^2_{\rm Pl}H^2_{\rm inf.}~~, 
\ee
the value of $H_{\rm inf.}/M_{\rm Pl}$ is suppressed by the factor 
$(\tilde{M}_{\rm Pl}/M_{\rm Pl})^2$. On the other hand, the inflationary Hubble scale is related to the Cosmic Microwave Background (CMB) tensor-to-scalar ratio $r$ as
\be \lb{Hr}
 \fracmm{H_{\rm inf.}}{M_{\rm Pl}}=1.06\cdot 10^{-4}\sqrt{r}~~. 
\ee
In turn, the $r$ is restricted by Planck (2018) measurements \cite{planck2018} as $r<0.064$ (with 95\% CL), which implies $H_{\rm inf.}< 6\cdot 10^{13}$ GeV. Therefore, the ratio $(\tilde{M}_{\rm Pl}/M_{\rm Pl})$ should be of the order $10^{-2}\div 10^{-3}\ll 1$ for viable inflation. This
justifies our setup in the Introduction (Sec.~1). We define the dimensionless parameter $\g$ as
$(\tilde{M}_{\rm Pl}/M_{\rm Pl})\equiv 10^{-3}/\g$, where $\g$ is of the order one.

In our numerical calculations, we have chosen the cutoff scale $\tilde{\L}=3$, so that the
restriction (\ref{bound}) is satisfied. Then Eqs.~(\ref{pot2}) and (\ref{crit}) imply
\be \lb{nump}
V_{\rm max}\tilde{M}_{\rm Pl}^{-4}=0.245 \quad {\rm and} \quad \s_{\rm cr.}=
\tilde{w}^{-1}(\phi_{\rm cr.}/M_{\rm Pl})=0.722~.
\ee
In turn, this yields the gravitino mass $m_{3/2}$ and the gravitino condensate mass $m_{\rm cond.}$ as follows: 
\be \lb{masses}
m_{3/2}=2.39\tilde{M}_{\rm Pl} \quad {\rm and} \quad m_{\rm cond.} =
\sqrt{8/11}\, m_{3/2} = 2.038  \tilde{M}_{\rm Pl}~.
\ee

We numerically studied the running of the slow inflationary parameters $\ve=\frac{1}{2}M^2_{\rm Pl}(V'/V)^2$ and $\eta=M^2_{\rm Pl}(V''/V)$ with respect to the inflaton field $\phi$ for the values of
the parameter $\g$ as $0.1$, $0.5$ and $1$, and found that $\ve$ is always under ${\cal O}(10^{-4})$ so that it can be ignored within the errors of Planck 2018 data. Then the value of the scalar index  $n_s=1-6\ve+2\eta = 0.9649\pm 0.0042$ (with 68\% CL) \cite{planck2018} can be reached with  $\eta=-0.0177$ at the horizon crossing by using the parameter $\tilde{w}$ of the order one. There are no additional constraints on the parameters $\g$ and $\tilde{w}$ from demanding the e-folding number
\be \lb{efn}
N_e = -\fracmm{1}{M^2_{\rm Pl}}\int^{\phi_{\rm end}}_{\phi_{\rm ini.}}\fracmm{V}{V'}d\phi
\ee
to be between 50 and 60, as is desired for viable inflation, when assigning the inflaton field 
$\phi/M_{\rm Pl}$ to run somewhere between $0$ and $5$ during inflation. The running of 
the slow-roll parameter $\eta$ is displayed in Fig.~2.

\begin{figure}[t]
\begin{center}
\vspace{1cm}
\includegraphics[width=15cm,height=8cm]{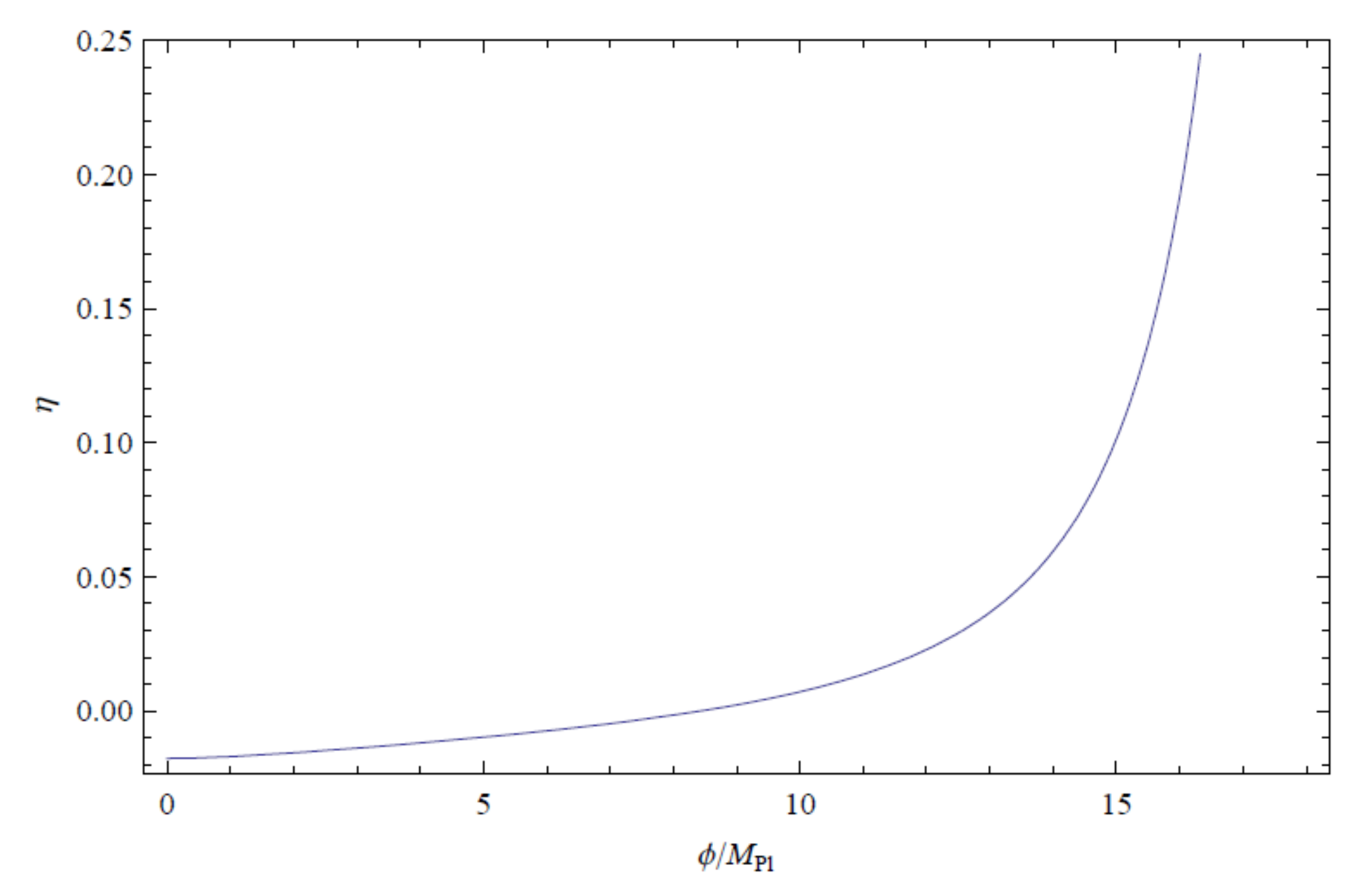}
\caption{The running of the slow-roll parameter $\eta$ for $\g=0.5$ and $\tilde{w}=13$. } 
\label{fig:2}
\end{center}
\end{figure}

In summary, our results qualitatively agree with those of Ref.~\cite{Ellis:2013zsa}, but quantitatively
allow considerably higher values of $\ve$ and $r$ up to the order ${\cal O}(10^{-4})$ contrary to
${\cal O}(10^{-8})$ of Ref.~\cite{Ellis:2013zsa}, with the Planckian values of the inflaton $\phi$ during inflation, contrary to its sub-Planckian values of   ${\cal O}(10^{-3})M_{\rm Pl}$ in Ref.~\cite{Ellis:2013zsa}. Hence, the inflationary scale  $H_{\rm inf.}$ can be as high as 
$10^{12}$ GeV versus $10^{10}$ GeV in Ref.~\cite{Ellis:2013zsa}.

\section{Adding the FI term}

In order to uplift the Minkowski vacuum to a de Sitter vacuum (dark energy) in our approach, we need an extra tool of
spontaneous SUSY breaking. In the  BI theory (without chiral matter) coupled to supergravity such tool can be provided by the (alternative) FI terms  \cite{Cribiori:2017laj,Kuzenko:2018jlz,Aldabergenov:2018nzd,Cribiori:2018dlc}  that do not require the gauged R-symmetry, unlike the standard FI term  \cite{Fayet:1974jb} whose extension to supergravity is severely restricted  \cite{Binetruy:2004hh}. 

The (abelian) gauge vector multiplet superfield $V$ can be decomposed  into a sum of the reduced gauge superfield ${\cal V}$ including the gauge field $A_{\mu}$, and the nilpotent gauge-invariant goldstino superfield ${\cal G}$ that contains only goldstino $\l$ and the auxiliary field $D$  \cite{Kuzenko:2018jlz},
\be \lb{dec}
V = {\cal V} + {\cal G}~,\quad {\cal G}^2 =0~.
\ee
The simplest examples of the goldstino superfield are given by 
\cite{Cribiori:2017laj,Kuzenko:2018jlz} 
\be \lb{fi1}
{\cal G}_1 = -4 \fracmm{W^2\bar{W}^2}{\mathcal{D}^2W^2
\bar{\mathcal{D}}^2\bar{W}^2}(\mathcal{D}W)
\ee
and
\be \lb{fi2}
{\cal G}_2 = -4 \fracmm{W^2\bar{W}^2}{(\mathcal{D}W)^3}~,
\ee
respectively, in terms of the standard $N=1$ gauge superfield strength
\begin{equation} \label{gss}
W_{\a} = -\fracmm{1}{4}  \left( \bar{\mathcal D}^2 - 4{\mathcal R}\right){\mathcal D}_{\a}V~,
\end{equation}
where $\mathcal R$ is the chiral scalar curvature superfield. The $W_{\a}$ obeys Bianchi identities
\begin{equation} \label{bis}
\bar{\mathcal D}_{\dt{\b}}W\low{\a}=0 \quad {\rm and} \quad \bar{\mathcal D}_{\dt{\a}}\bar{W}^{\dt{\a}}
\equiv \bar{{\mathcal D}}\bar{W}={\mathcal D}^{\a}W_{\a}\equiv {\mathcal D}W~.
\end{equation}
The field components are given by $\left.W_{\a}\right|=\l_{\a}$,  $\left.{\mathcal D}W\right|=-2D$,
and  $\left.{\mathcal D}_{(\a}W_{\b)}\right|=i(\s^{ab})_{\a\b}F_{ab}+\ldots$ The difference between the superfields  ${\cal G}_1$ and ${\cal G}_2$ is only in the gauge sector, while it is not essential for our purposes here.

The extra FI term with the coupling constant $\x\neq 0$ is given by
\be \label{fit}
S_{\rm FI} = \x \int d^4x d^4\theta E {\cal G}~,
\ee
where $E$ is the supervielbein (super)determinant \cite{Wess:1992cp}.  This FI term is manifestly SUSY- and gauge-invariant, does {\it not\/}  include the higher spacetime derivatives of the field components, but leads to the inverse powers of the auxiliary field $D$ (up to the fourth order) in the non-scalar sector of the theory.~\footnote{The limit $\xi\to 0$ does not lead to a well defined theory, so that $\VEV{D}=\xi$ must be non-vanishing.}  Integrating out the auxiliary field $D$ leads to the {\it positive} contribution to the cosmological constant
\begin{equation} \label{cc}
V_{\x}= \ha \xi^2 >0~.
\end{equation}
Matching $V_{\x}$ with the observed cosmological constant allows us to include a viable description of the dark energy into our approach. The phenomenological values of the cosmological constant and the related contribution $(\xi)$ to the vacuum expectation value of the auxiliary field  $D$ are tiny, so that they do not affect our considerations of the high-scale SUSY breaking in the previous Sections.

The nilpotent Goldstino superfield ${\cal G}$ introduced above is composed of the usual (standard) superfields and, hence, is very different from the intrinsically nilpotent goldstino superfield introduced in 
Refs.~\cite{Rocek:1978nb,Lindstrom:1979kq,Komargodski:2009rz,Antoniadis:2014oya}. 

As the FI term affects the quartic and higher-order terms with respect to the gauge field and its fermionic (spin-1/2) superpartner, back reaction of the FI term on the effective action should be examined (work in progress). It should be done together with quantum renormalization of those terms and, perhaps, requires a field-dependent FI parameter $\xi$.
 The $D$-type scalar potential and the associated dark energy are expected to be unaffected because of cancellation of (perturbative) quartic and quadratic (ultra-violet) divergences due to supersymmetry of the action.
\vglue.2in

\section{Conclusion}

The gravitino condensate can be considered as a viable candidate for inflaton in supergravity,
when assuming the effective (quantum) gravity scale to be close to the (super)GUT scale that is 
also close to the SUSY breaking scale in our approach, with all scales close to $10^{15}$ GeV. Actually, in this scenario we have the hyper-GUT where {\it all} fundamental interactions merge, including gravity. At the same time, it is the weak point of our calculations, because we ignored (other) quantum gravity corrections.

The inflationary (Hubble) scale is well below the GUT scale, and can be as large as $10^{12}$ GeV. The gravitino mass is above the inflationary scale, so that there is no gravitino overproduction problem in the early Universe. The constraints from proton decay and the Big Bang Nucleosynthesis (BBN) are very weak because of high-scale SUSY. Then SUSY is not a solution to the hierarchy problem with respect to the electro-weak scale. This is similar to the setup studied in Refs.~\cite{Dudas:2017rpa,Dudas:2017kfz}. Our scenario is consistent with the known Higgs mass of about $125$ GeV after taking into account the extreme possible values of the gaugino mixing parameter $\tan\beta$ in the context of SUSY extensions of the Standard Model \cite{hssusy}.

As regards reheating after inflation, the inflaton (gravitino condensate) field decays into other matter and radiation, which is highly model-dependent, as usual. Unlike Ref.~\cite{Addazi:2017ulg}, the inflaton as the gravitino condensate cannot decay into gravitinos because Eq.~(\ref{masses})
leads to the kinematical constraint $2m_{3/2}>m_{\rm cond.}$. It also implies that gravitino cannot
be a dark matter particle in this scenario. A detailed study of reheating requires the knowledge of couplings of gravitino and gravitino condensate to the Standard Model particles, which is  beyond the scope of this paper.

\newpage

\section*{Acknowledgements}

SVK was supported by the World Premier International Research Center Initiative (WPI), MEXT, Japan, and  the Competitiveness Enhancement Program of Tomsk Polytechnic University in Russia. SVK is grateful to Albert Einstein Institute for Gravitational Physics of Max Planck Society in Golm, Germany, for the hospitality extended to him during the preparation of this paper, and Andrea Addazi, Dmitri Bykov, Gia Dvali, Maxim Khlopov, Sergey Kuzenko and Kai Schmitz for discussions and correspondence. 




\end{document}